\documentclass{appolb}
\usepackage{amsmath,amssymb}
\usepackage{graphicx}
\usepackage{color}
\def\permille{\ensuremath{{}^\text{o}\mkern-5mu/\mkern-3mu_\text{oo}}}

\begin{document}
\title{Status of the superweak extension 
of the standard model and muon $g-2$%
\thanks{Presented at Matter to the Deepest 2023 and in part at the
23rd Hellenic School and Workshop on Elementary Particle Physics
international conferences.  \\
This research was supported by the Excellence Programme of the Hungarian
Ministry of Culture and Innovation under contract TKP2021-NKTA-64. \\
The author is grateful to  the Galileo Galilei Institute for
Theoretical Physics for the hospitality and the INFN for partial
support during the completion of this work.
}%
}
\author{Zolt\'an Tr\'ocs\'anyi
\address{Institute for Theoretical Physics,
ELTE E\"otv\"os Lor\'and University,
P\'azm\'any P\'eter s\'et\'any 1/A, 1117 Budapest, Hungary\\
Institute of Physics, Faculty of Science and Technology,
University of Debrecen,
Bem t\'er 18/A, 4026 Debrecen, Hungary}
}
\maketitle
\begin{abstract}
The super-weak force is a minimal, anomaly-free U(1) extension of the
standard model, designed to explain the origin of (i) neutrino masses
and mixing matrix elements, (ii) dark matter, (iii) cosmic inflation,
(iv) stabilization of the electroweak vacuum and (v) leptogenesis. In
this talk we discuss the phenomenological status of the model and 
provide viable scenarios for the physics of the items in this list.
\end{abstract}
  
\section{Introduction}
The standard model (SM) of particle interactions has reached a mature and
robust status. We do not have doubt that it correctly describes the
scattering processes at colliders \cite{CMS-SMpublic}. At the same time,
the experiments at the LHC have not yet found any sign of new physics
yet, although exciting deviations from the SM predictions keep appearing,
such as a recent bump at 146\,GeV center-of-mass energy in the
process
${\rm pp} \to X \textrm{(= new Higgs boson)} \to {\rm e}^\pm\mu^\mp$
\cite{CMS:2023pte}. However, none of these have reached 5\,$\sigma$
significance, and we do not discuss those further in this presentation.

In spite of the success of the SM, we are also certain that it cannot be
the final theory of the microworld.  There are pressing questions in
the cosmic and intensity frontiers \cite{ParticleDataGroup:2022pth}:
(i) What does the non-baryonic dark matter (DM) consist of?
(ii) What gives masses to the neutrinos?
(iii) What is the origin of the matter--anti-matter asymmetry?
(iv) How can we explain epochs of accelerated expansion of the Universe?
These established observations require physics beyond the standard model
(BSM), but the lack of further discoveries do not suggest a rich BSM
physics at energies accessible at the LHC. The main theme of this talk is
to discuss whether the observations (i--iv) can be explained by a
consistent, simple extension of the SM.

\section{Status of the muon anomalous magnetic moment}

There is one long-standing anomaly that may also require BSM explanation
and received new momentum recently. The Fermilab Muon $g-2$ Experiment has
measured the anomalous magnetic moment $a_\mu = \frac{g-2}{2}$ of the
anti-muon with an unprecedented precision and in agreement with previous
measurements, leading to a new world average $a_\mu = 116\,592
\,059(22)\cdot 10^{-11}$ \cite{Muong-2:2023cdq}.  This result differs
from the SM prediction of the Muon $g-2$ Theory Initiative
\cite{Aoyama:2020ynm} by 5.0\,$\sigma$. This theory prediction extracts
the leading-order contribution to the hadronic vacuum polarization (HVP) 
from measurments of the total hadronic cross section at low energies
using the optical theorem.
Such measurements exhibit tension among the results of different
experiments in the energy range $\sqrt{s} \in [600,880]$\,MeV, which
gives more than half of the total contribution of HVP to $a_\mu$. In
fact, the cross section for the process $e^+ e^- \to \pi^+ \pi^-$ from
the recent CMD3 experiment \cite{CMD-3:2023alj}
disagrees with all other $e^+e^-$ data, and hence with the old world
average at 4.4\,$\sigma$. If confirmed, this new measurement would mean a
15 unit increase in $10^{10} a_\mu$ as compared to the earlier prediction,
which would mean compatibility with the prediction
obtained by a lattice computation of HVP by the BMW collaboration 
\cite{Borsanyi:2020mff}, also recently confirmed partially by
independent lattice calculations. 

As argued, we are certain about the existence of BSM ohysics, and it is
likely to contribute to the $a_\mu^{\rm (SM)}$ value by a positive shift.
Thus presently the main question concerning the value of $\Delta a_\mu =
a_\mu^{\rm (exp)} - a_\mu^{\rm (SM)}$ is the expected size of the new
physics contribution: whether it is ``large'' (accounting for the 
$5\,\sigma$ difference), or ``small'' (meaning insignificant as predicted
on the lattice). The experimental result appears robust, only its
uncertainty will reduce further by a factor of two when all Fermilab data
will be analysed. The main task is to resolve the discrepancy between the
SM theory predictions. There are ongoing efforts to clarify
the current theoretical situation \cite{Colangelo:2022jxc}. Until
reaching a conclusion in this respect everything else is a mere
speculation.

In general, the BSM contirbution to $a_\mu$ is proportional to the square
of the muon mass, and therefore, on dimensional considerations inversely
proportional to the square of the mass of the BSM particle 
\cite{Czarnecki:2001pv}, 
\begin{equation}
\Delta a_\mu^{\rm BSM} = C_{\rm BSM} \frac{m_\mu^2}{M_{\rm BSM}^2}
\,.
\end{equation}
As the same particle also has quantum corrections to the mass of the
muon, in order that this loop correction not to be too large, the
coefficient can at most be of O(1). Hence, a large BSM contribution to
$a_\mu$ can only be explained  by rather small masses and/or large
couplings of the BSM particle and its enhanced chirality flips%
\footnote{The QFT operator corresponding to $a_\mu$ connects left and
right chirality muons.},
so using the $R$-ratio prediction for HVP, we find an
upper limit for the mass of the BSM particle,
\begin{equation}
\Delta a_\mu^{\rm BSM}
\lesssim {\rm O}(1) \frac{m_\mu^2}{M_{\rm BSM}^2}
\Rightarrow M_{\rm BSM} \lesssim 2\,{\rm TeV}
\,,
\end{equation}
which often leads to conflicts with lower limits from LHC and dark matter
experiments.

Indeed, an extensive study of single-, two- and three-field extensions
of the SM that can explain the large value of $\Delta a_\mu^{\rm BSM}$
was carried out in Ref.~\cite{Athron:2021iuf} to check which
is still allowed. Most of these extensions are already excluded. The few
remaining possibilities are incomplete three-field models with two
fermions and a scalar or one fermion and two scalars. The Minimal
Supersymmetric extension of the SM is also still a vialble explanation of
a large $\Delta a_\mu^{\rm BSM}$ if the lightest supersymmetric particle
is Bino or Wino like \cite{Endo:2021zal,Iwamoto:2021aaf}. In the latter
case however, the dark matter abundance requires additional DM candidates.

As muon  flip enhancements are related to the mass generation mechanism
for the muon, the measurement of the Higgs–muon coupling at LHC or FCC
can (and hopefully will) provide further tests. However, as argued the
resolution of the tension between the theory predictions has 
priority where the proposed MUoNe \cite{Banerjee:2020tdt}
experiment should play a decisive role. If the small $\Delta a_\mu^{\rm
BSM}$ as predicted by the lattice computations becomes confirmed, then
the new physics contirbution to $a_\mu$ is smaller than the electorweak
correction. In this case the precise measurement of $a_\mu$ may constrain
the parameter space of the model describeing BSM physics, but it is
unlikely to exclude a model that is compatible with Electroweak
Precision Observables (EWPOs).

\section{Going beyond the standard model}

Broadly speaking there are three classes of extensions of the SM, each
with their own strengths and weaknesses. The effective field theory (EFT)
approach, such as the SMEFT is completely general, but also higly complex
with its 2499 dimension six operators (and even more higher dimensional
ones). It focuses on new physics at high energy scales. On the other end
of the spectrum simplified models, such as dark photon, extended scalar
sector or right-handed neutrinos provide a reasonably easily accessible
phenomenology.  However, these cannot explain all BSM phenomena
simultaneously as they focus on specific aspects of new physics.

A less ambitious approach than SMEFT is the SuperWeak extension of the
SM (SWSM) that belongs to the third class of models together with for
instance the supersymmetric extensions of the SM. The SWMS is a
phenomenological, ultraviolet complete extension designed such that it
could explain all firmly observed BSM phenomena,
but not more \cite{Trocsanyi:2018bkm}. In this model the field content of
the SM is supplemented by three right-handed SM sterile neutrinos
$\nu_{{\rm R},1}$, $\nu_{{\rm R},2}$, $\nu_{{\rm R},3}$, and a complex
scalar $\chi$ whose non-vanishing vacuum expectation value (VEV) $w$
breaks the new $U(1)_z$ symmetry that is added to the SM symmetry group
$G_{\rm SM}$. The model contains all dimension four renormalizible
operators allowed by $G_{\rm SM}\otimes U(1)_z$. The new charges belonging
to the new gauge interaction are determined by cancellation of the gauge
and gravity anomalies. Up to two unkown $z$-charges. One of these is set
by the gauge invariant Yukawa terms needed for neutrino mass generation,
while the second one can be set at wish, defining the normalization of
the new gauge coupling $g_z$.

\section{Superweak extension of the standard model}

The SWSM contains three neutral gauge bosons: in addition to the SM
fields $W_3^\mu$ and $B^\mu$, there is also an Abelian field
$B^{\prime\mu}$. These fields mix into mass eigenstates $A^\mu$,
$Z^{0\mu}$ and $Z^{\prime\mu}$ by two rotations,
\begin{equation}
\begin{pmatrix}
B^\mu \\
W_3^\mu \\
B^{\prime\mu}
\end{pmatrix}
= \begin{pmatrix}
c_W & -s_W & 0
\\
s_W & ~~ c_W & 0
\\
0 & 0 & 1
\end{pmatrix}
\begin{pmatrix}
1 & 0 & 0
\\
0 & c_Z & -s_Z
\\
0 & s_Z & ~~ c_Z
\end{pmatrix}
\begin{pmatrix}
A^\mu \\
Z^\mu \\
Z^{\prime\mu}
\end{pmatrix}
\end{equation}
where $c_X=\cos\theta_X$ and $s_X = \sin\theta_X$, with $X = W$ for the
weak mixing angle and $X=Z$ for the new $Z-Z'$ mixing. The latter one can be
given in a simple, but implicit form in terms of two effective couplings
$\kappa$ and $\tau$, which are functions of the Lagrangian couplings
\cite{Trocsanyi:2018bkm}, as
\begin{equation}
\tan(2\theta_Z) = -\frac{2\kappa}{1-\kappa^2-\tau^2}
\,.
\end{equation}
The tree-level masses of the neutral gauge bosons can be expressed with
the mass $M_W = \frac12 g_{\rm L} v$ of the $W$ bosons, $c_W$ and
$\kappa$, $\tau$. While these formulas are somewhat cumbersome, there
exist a nice, compact generalization of the SM mass formula $M_W = c_W
M_Z$ as follows \cite{Peli:2023fyb}:
\begin{equation}
\frac{M_W^2}{c_W^2} = c_Z^2 M_{Z^0}^2 + s_Z^2 M_{Z'}^2
\label{eq:MW}
\,.
\end{equation}

The scalar potential of the Brout-Englert-Higgs field $\phi$ and the new
scalar $\chi$ is
\begin{equation}
\label{eq:V_phichi}
    V(\phi,\chi) = V_0 - \mu_\phi^2 |\phi|^2 - \mu_\chi^2 |\chi|^2
    +\lambda_\phi |\phi|^4 + \lambda_\chi |\chi|^4
    +\lambda |\phi|^2|\chi|^2
\subset  -\mathcal{L}
\end{equation}
where $|\phi|^2 =|\phi^+|^2 + |\phi^0|^2$.
In the $R_\xi$ gauge we parametrize the scalar fields after
spontaneous symmetry breaking as
\begin{equation}
\phi =\frac{1}{\sqrt{2}}\binom{-{\rm i} \sqrt{2}\sigma^+}{v+h'+ {\rm i}\sigma_\phi}
\,,\quad
\chi = \frac{1}{\sqrt{2}}(w + s' + {\rm i}\sigma_\chi)
\end{equation}
where $v$ and $w$ denote the VEVs of the scalar fields. The scalar field
mass eigenstates $h$ and $s$ are 
\begin{equation}
\binom{h'}{s'}
= \begin{pmatrix}
c_S & -s_S
\\
s_S & ~~ c_S
\end{pmatrix}
\binom{h}{s}
\end{equation}
where $\theta_S$ is the scalar mixing angle, given implicitly by
\begin{equation}
\tan(2\theta_S) = \frac{\lambda v w}{(\lambda_\chi w^2 - \lambda_\phi
v^2)}
\,.
\end{equation}

The SWSM has five new parameters besides the new couplings in the Yukawa
sector. In the Lagrangian these are
new gauge couplings $g_z$ and $g_{yz}$, the latter characterizing the
mixing of the two $U(1)$ fields. Furthermore, in the scalar sector out of 
the five couplings two are constrained by the known $v$ and $M_h$,
leaving three unknown. 
Alternatively, in the gauge sector we have the effective couplings
$\kappa$, $\tau$ and in the scalar sector $w$, $\lambda_\chi$ and
$\lambda$, or the phenomenologically more accessible mixing angles
$\theta_Z$ and $\theta_S$, new boson masses  $M_{Z'}$ and $M_S$, plus
the scalar mixing coupling $\lambda$. The different sets have different
merits.

In the fermion sector of the SWSM the masses of the neutrinos
are generated after SSB by the new Yukawa terms
\begin{equation}
\frac{1}{2}\overline{\nu_R}\textbf{Y}_N(\nu_R)^c\chi
+ \overline{\nu_R}\textbf{Y}_\nu \varepsilon_{ab}L_{La}\phi_b + {\rm h.c.}
\subset -\mathcal{L}
\,,
\end{equation}
which leads to a $6\times 6$ mass matrix 
\begin{equation}
\textbf{M}' =
\begin{pmatrix}
\textbf{0}_3 & \textbf{M}_D^T \\
\textbf{M}_D & \textbf{M}_N
\end{pmatrix}
\,,
\end{equation}
with
\begin{equation}
\textbf{M}_N = \frac{w}{\sqrt{2}}\textbf{Y}_N
\,,\quad
\textbf{M}_D = \frac{v}{\sqrt{2}}\textbf{Y}_\nu
\,.
\end{equation}
As the left- and right-handed neutrinos have the same quantum numbers,
they may mix, leading to a type-I see-saw masses of the active and
sterile neutrinos. It is important that both the Dirac and Majorana
mass-terms $\textbf{M}'$ appear already at tree level. It was shown in
Ref.~\cite{Iwamoto:2021wko} that the quantum corrections to the masses of
the active neutrinos remain perturbatively small over most of the
parameter space.

\section{Expected consequences of the SWSM}

The take-home messages of this talk can be summarized as follows:
\begin{enumerate}
\itemsep=-2pt
\item
Dirac and Majorana neutrino mass terms are generated by the SSB of the
scalar fields, providing the origin of neutrino masses and oscillations
\cite{Iwamoto:2021wko,Karkkainen:2021tbh}.
\item
The lightest new particle is a natural and viable candidate for WIMP
dark matter if it is sufficiently stable \cite{Iwamoto:2021fup}.
\item
Diagonalization of neutrino mass terms leads to the PMNS matrix, 
which in turn can be the source of lepto-baryogenesis. This is being
explored in ongoing research \cite{Seller:2023xkn}.
\item
The second scalar together with the established BEH field can stabilize
the vacuum and be related to the accelerated expansion now and inflation
in the early universe \cite{Peli:2019vtp,Peli:2022ybi}.
\end{enumerate}
In the rest of this contribution we shall discuss briefly the status of
some of these consequences of the model.

There are two important questions to answer if we want to explore if
Nature realizes this model: (i) Is there a non-empty region of the
parameter space where the listed promises are fulfilled? (ii) Can we
predict any new phenomenon observable by present or future experiments?
Of course, an important test of the SWSM will be the observation of a
$Z'$ gauge boson and a new scalar $S$ in the allowed region of the
parameter space.

\section{Dark matter candidate}

We have evidence that DM exists, but it is based solely on its
gravitational effect. While DM can have cosmological origin, in particle
physics we assume naturally that it is a new kind of particle. The only
chance to observe such a particle is the use of detectors of ordinary
matter, hence the interaction of DM with SM particle, which requires a
portal. The natural portal in the SWSM is the $Z'$, with the lightest
sterile neutrino as DM candidate. The latter has to be sufficiently
stable, which requires a negligible mixing between the active and the
sterile neutrinos.

\begin{figure}[htb]
\centerline{%
\includegraphics[width=12.5cm]{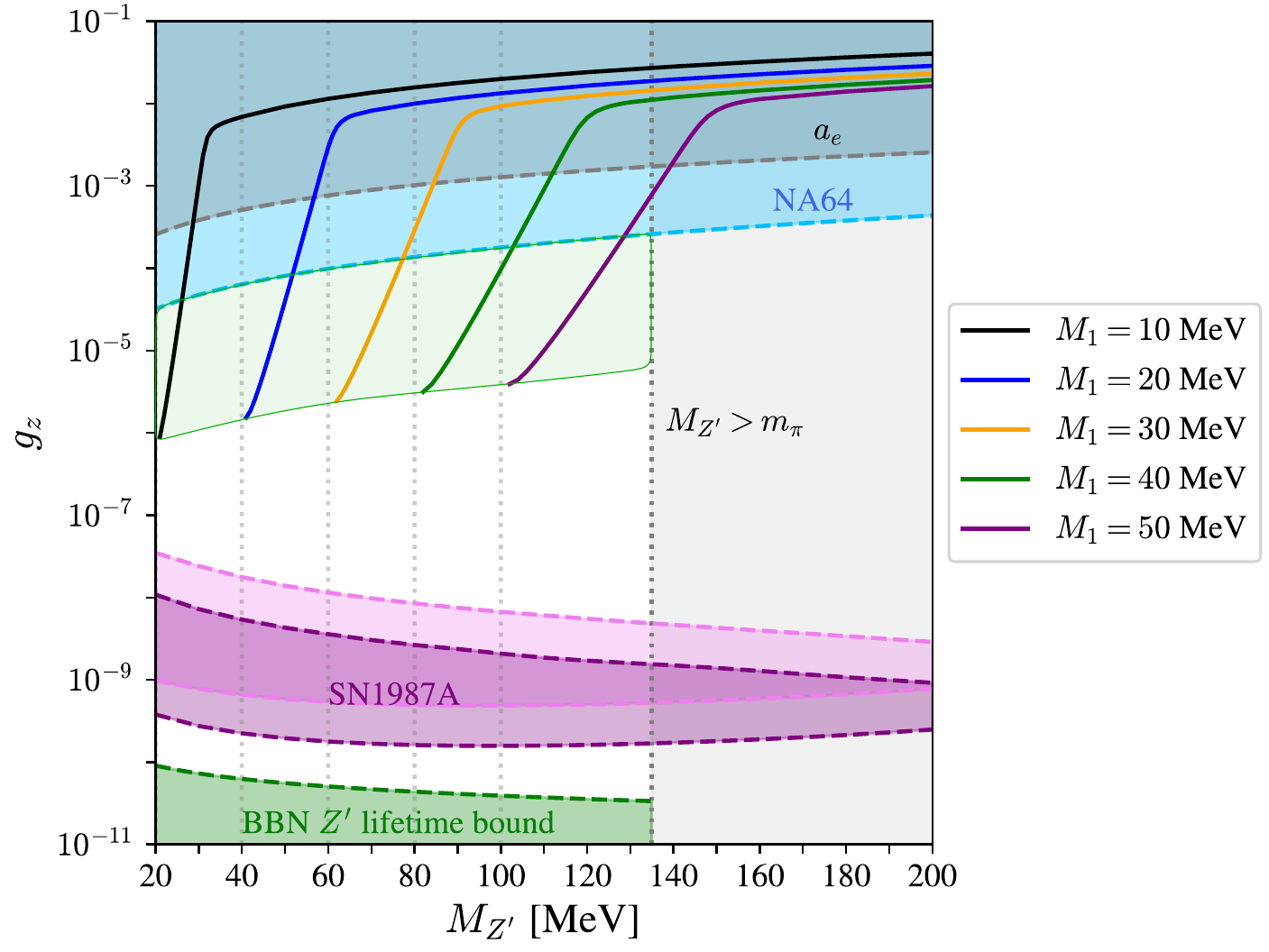}}
\caption{Parameter space for the freeze-out scenario of dark matter
production in the SWSM. The meanings of lines and shaded regions are
explained in the main text.}
\label{fig:freezeout}
\end{figure}

It was found in Ref.~\cite{Iwamoto:2021fup} that the SWSM can provide the
correct relic abundance of DM both with freeze-in and freeze-out
mechanisms. The former requires very small portal couplings, hence it is
more difficult to verify or exclude experimentally.
Fig.~\ref{fig:freezeout} shows the parameter space in the $g_z-M_{Z'}$
plane in the freeze-out case. Each line corresponds to a fixed value of
the DM neutrino mass providing the correct DM relic density, while the
shaded regions are excluded by the experimental results for the
anomalous magnetic moment of the electron and the direct searches for
dark photon by the NA64 experiment \cite{NA64:2017vtt}.

We see that the correct DM relic density requires the new gauge coupling
$g_z$ too large, excluded by other measurements (close to horizontal
slopes), except for the case of {\em resonant annihilation} of the
lightest sterile neutrinos (steep slopes), i.e.~when $M_{Z'} \approx 2
M_{\nu_{R1}}$. The smaller portal coupling the earlier freeze-out
time, and hence the larger DM abundance. Resonant annihilation is
needed in order to enhance the probability of DM depletion without the
increase of $g_z$.

In addition to particle physics constraints on the possible values of
$g_z$, such as the anomalous magnetic moment of the charged leptons,
direct searches for dark photons and beam dump experiments constraining
the possible $Z'$ masses, there are also cosmological measurements that
limit the possible value of the protal coupling. Big Bang Nucleosynthesis
(BBN) has a fairly robust experimental support. The theory of BBN does
not allow for a significant contribution to the creation of light mesons
during BBN. As the $Z'$ in the SWSM interacts with the quarks, its mass
should be below the pion threshold, shown by the vertical grey
exclusion region in the figure. Other cosmological bounds were estimated
not to influence the parameter space relevant for the freeze-out scenario.

\section{Estimates of phase transition temperatures}

\begin{figure}[htb]
\centerline{%
\includegraphics[width=12.5cm]{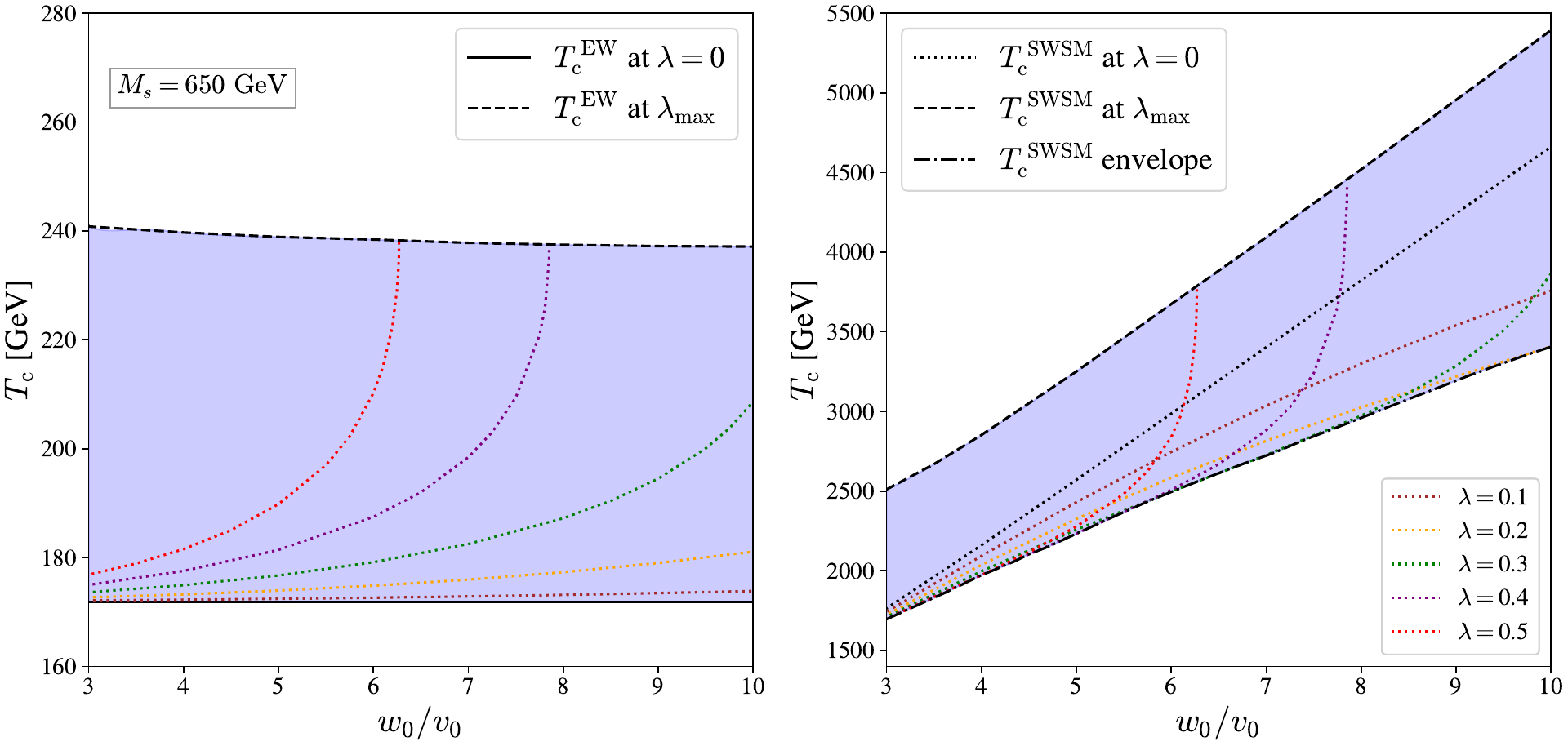}}
\caption{Electroweak and superweak phase transition temperatures
estimated using high-temperature perturbation theory at one-loop order
in the SWSM at fixed value of the new scalar mass. The meanings of
lines and shaded regions are explained in the main text.}
\label{fig:Tc}
\end{figure}

There is accumulating evidence that the $CP$-violating phase in the
lepton sector can be much larger than in the quark sector \cite{T2K:2023smv},
which means that leptogenesis may provide an explanation for the observed
baryon asymmetry, if it is not washed out by particle processes. That
requires an epoch in the Universe with heavy neutral leptons (the
massive sterile neutrinos in the SWSM) and sphaleron process
\cite{Kuzmin:1985mm} allowed. The former gains mass already from
the $w$ VEV through the Majorana-type Yukawa term in the Lagrangian%
\footnote{The smallness of the $g_z$ coupling in the SWSM implies that
the thermal mass of the HNL is small compared to its mass at SSB.},
while the latter stops when the sphaleron rate drops below the Hubble
rate near the electroweak phase transition \cite{DOnofrio:2014rug}.
Hence one has to estimate the critical temperatures of the superweak
and electroweak phase transitions, which has been performed in
Ref.~\cite{Seller:2023xkn} (see also K.~Seller's contribution
\cite{Seller:2023juo}).  Fig.~\ref{fig:Tc} shows the critical
temperatures as a function of the ratio of the VEVs at a selected value
of the mass of the new scalar boson. The shaded region is swept by the
lines of constant scalar mixing couplings (shown at several selected
values) in the region $\lambda \in [0,\lambda_{\max}(w)]$ where
$\lambda_{\max}(w)$ is the value of its largest possible value such
that a parametrization of the effective potential exists in terms of
real Lagrangian parameters.  The figure provides a benchmark case when
the superweak phase transion happens well before the electroweak one,
giving ample opportunity for leptogenesis.

\section{Scalar sector constraints}

\begin{figure}[htb]
\centerline{%
\includegraphics[width=6.2cm]{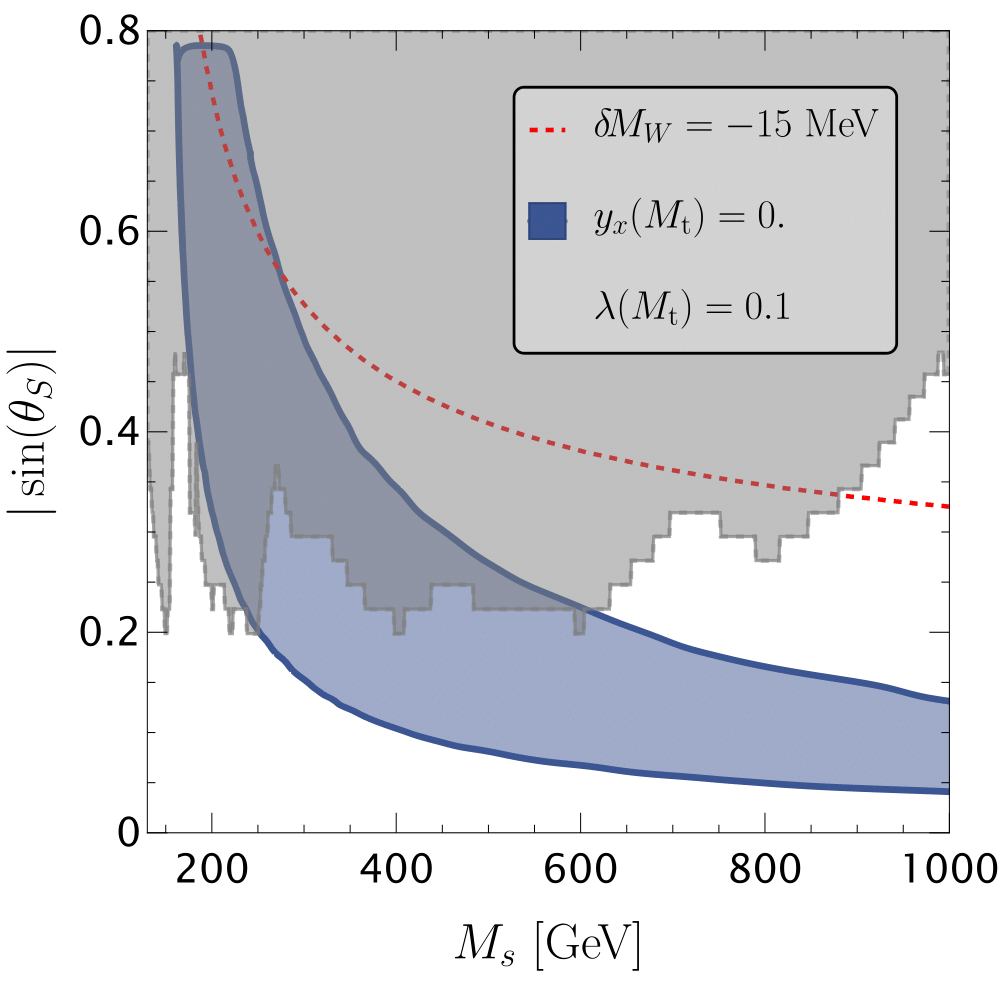}
\hfill
\includegraphics[width=6.2cm]{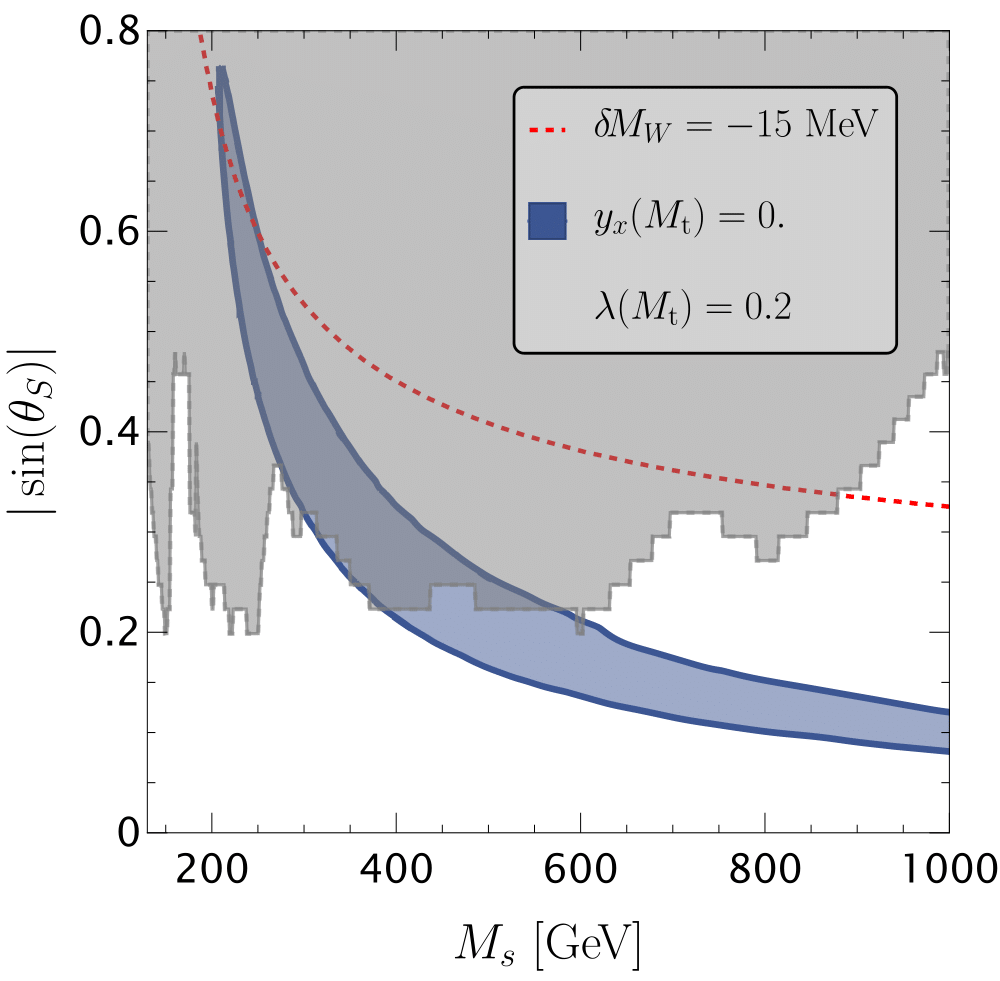}}
\caption{Excluded region of the scalar mixing angle as a function of the
new scalar mass. The meanings of lines and shaded regions
are explained in the main text.}
\label{fig:scalarlimits}
\end{figure}

Collider experiments have always been searching for new scalars. The
exclusion limits for the value of the scalar mixing angle as a function
of the new scalar mass $M_S$ was measured at the LHC in the mass range
above the Higgs mass up to 1\,TeV as shown in Fig.~\ref{fig:scalarlimits}
by the the shaded region, which still leaves ample parameter space for
the SWSM. The banana-shape strips correspond to the region where the
SWSM vacuum remains perturbatively stable up to the Planck mass computed
at two-loop accuracy in perturbation theory at vanishing Majorana-type
Yukawa couplings $y_x$, considered equal in this example. Incresing the
latter, the region becomes narrower and vanishes for values slightly
above $y_x = 0.8$. The region above the dashed line is excluded by the
precision measurements for the mass of the $W$ bosons. The experimental
precision of the latter has reached of about one per myriad
\cite{ParticleDataGroup:2022pth}, making it an important Electroweak
Precision Observable (EWPO) as we discuss next. 

\section{Prediction for the W boson mass in the SWSM}

The new gauge boson couples to all particles, hence contributes to all
EWPO quantities. There are strong limits on EWPOs set by former and
current high-energy scattering experiments, which must be respected by
the predictions in the SWSM. An example is the mass of the $W$ boson that
can be determined from the decay width of the muon \cite{Sirlin:1980nh},
\begin{equation}
M_W^2 = \frac{M_Z^2}{2}
\left(1+ \sqrt{1- \frac{4 \pi\alpha\Big/\Big(\sqrt{2}G_F\Big)}
{M_Z^2}\frac{1}{1-\Delta r_{\rm SM}}} \right)
\end{equation}
where $\Delta r_{\rm SM}$, which collects the quantum corrections, is
already known completely at two- and partially at three loops in the SM.
The SWSM introduces three types of corrections to this formula, exhibited
in read:
\begin{equation}
\begin{split}
M_W^2 &= \frac{{\color{red} c^2_Z} M_Z^2
{\color{red} + s^2_Z M_{Z'}^2}}{2}
\\&\times
\left(1+ \sqrt{1- \frac{4 \pi\alpha\Big/\Big(\sqrt{2}G_F\Big)}
{{\color{red} c^2_Z} M_Z^2 {\color{red} + s^2_Z M_{Z'}^2}}
\frac{1}{1-\Delta r_{\rm SM}
{\color{red} - \Big(\Delta r^{(1)}_{\rm BSM} + \Delta r^{(2)}_{\rm BSM}\Big)}}}
\right).
\end{split}
\end{equation}
We recognize the tree-level corrections of Eqn.~\eqref{eq:MW}, and two classes
of loop corrections: (i) $\Delta r^{(1)}_{\rm BSM}$ collects the same
type of diagrammatic corrections as $\Delta r_{\rm SM}$ does in the SM
but with new particles in the loop, while (ii) $\Delta r^{(2)}_{\rm
BSM}$ contains the quantum corrections to the mixing parameter $s_Z$.
This second type of correction is often neglected in the
literature (e.g.~in the public code of Ref.~\cite{Athron:2014yba}).
Z.~P\'eli's contribution to these proceedings \cite{Peli:2023mttd} gives
a detailed account where in the parameter space such an approximation
may lead to insufficient accuracy in the theoretical prediction for $M_W$.

\section{Conclusions}

In this contribution we have presented the current phenomenological
status of the superweak extension of the standard model of particle
interactions. We argued that established observations at the three
frontiers of particle physics require physics beyond the standard model,
but do not suggest rich beyond the standard model physics. The $U(1)_z$
superweak extension has the potential of explaining all known results
beyond the standard model:
(i) Neutrino masses are generated by spontaneous symmetry breaking at
tree level. The one-loop corrections to the tree-level neutrino mass
matrix is known, and they are small (below 1\,$\permille$) in the
parameter space relevant in the superweak extension. 
(ii) The lightest sterile neutrino is a candidate
DM particle in the [10,50] MeV mass range for freeze-out mechanism with
resonant enhancement, which provides a prediction of the superweak
phenomenology, namely an approximate mass relation between vector boson
and lightest sterile neutrino.
(iii) In the scalar sector we find non-empty parameter space for a new
scalar that is heavier than the Higgs boson.
(iv) Contributions to electroweak precision observables, such as lepton
$g-2$ or $W$ boson mass, are negligible in the superweak region where
freeze-out dark matter scenario is realistic and a systematic
exploration of the parameter space is an ongoing project.

\end{document}